\documentclass[referee]{raa}            
\usepackage[a4paper, margin=1in]{geometry} 
\usepackage{graphicx,times}             
\usepackage{natbib}
\usepackage{amssymb,amsmath}
\usepackage{subcaption}
\bibpunct{(}{)}{;}{a}{}{,}

\usepackage[pagebackref=true]{hyperref}

\begin{document}

\title{The Short-spacing Interferometer Array for Global 21-cm Signal Detection (SIGMA): Design of the Antennas and Layout}


   \volnopage{Vol.0 (20xx) No.0, 000--000}      
   \setcounter{page}{1}          

\author{Feiyu Zhao
   \inst{1,2}
   \and Quan Guo
   \inst{1,3} 
   \and Qian Zheng
   \inst{1,3} 
   \and Ruxi Liang
   \inst{1}
   \and Pengfei Zhang
   \inst{4}
   \and Yajun Wu
   \inst{1}
   \and Junhua Gu
   \inst{5} 
   \and Zhao Yang
   \inst{4}
   \and Yun Yu
   \inst{1}
   \and Yan Huang
   \inst{5}
   \and Tianyang Liu
   \inst{1,2}
}

\institute{
   Shanghai Astronomical Observatory, Chinese Academy of Sciences, 80 Nandan Road, Shanghai 200030, China; {\it guoquan@shao.ac.cn, qzheng@shao.ac.cn, jhgu@nao.cas.cn}\\
   \and
   University of Chinese Academy of Sciences, No.1 Yanqihu East Road, Beijing 101408, China\\
   \and
   Key Laboratory of Radio Astronomy and Technology, Chinese Academy of Sciences, 20A Datun Road, Beijing 100101, China\\
   \and
   Xidian University and National Key Laboratory of Radar Detection and Sensing, No. 2, South Taibai Road, Xi'an 710071, China\\
   \and
   National Astronomical Observatories, Chinese Academy of Sciences, 20A Datun Road, Beijing 100101, China\\
\vs\no
   {\small Received 20xx month day; accepted 20xx month day}}

\abstract{ Numerous experiments have been designed to investigate the Cosmic Dawn (CD) and Epoch of Reionization (EoR) by examining redshifted 21-cm emissions from neutral hydrogen. Detecting the global spectrum of redshifted 21-cm signals is typically achieved through single-antenna experiments. However, this global 21-cm signal is deeply embedded in foreground emissions, which are about four orders of magnitude stronger. Extracting this faint signal is a significant challenge, requiring highly precise instrumental calibration. Additionally, accurately modelling receiver noise in single-antenna experiments is inherently complex. An alternative approach using a short-spacing interferometer is expected to alleviate these difficulties because the noise in different receivers is uncorrelated and averages to zero upon cross-correlation. The Short-spacing Interferometer array for Global 21-cm Signal detection (SIGMA) is an upcoming experiment aimed at detecting the global CD/EoR signal using this approach. We describe the SIGMA system with a focus on optimal antenna design and layout, and propose a framework to address cross-talk between antennas in future calibrations. The SIGMA system is intended to serve as a prototype to gain a better understanding of the system's instrumental effects and to optimize its performance further.
}

\authorrunning{F. Zhao et al.} 
\titlerunning{Design of the SIGMA} 

   \maketitle

%
%
\section{Introduction}
The redshifted 21-cm signal of neutral hydrogen from the Cosmic Dawn (CD) and Epoch of Reionization (EoR) is detectable in the low-frequency radio band, ranging from 50 to 200 MHz. Theoretical models suggest that this signal is extremely weak, overshadowed by foreground emissions that are 4-5 orders of magnitude higher (\citealt{furlanetto2006cosmology}). Several radio telescopes have been constructed with the detection of the CD/EoR signal as a primary scientific goal. Methods for detecting CD/EoR include: (a) statistical analysis of CD/EoR structures using radio interferometers; (b) tomographic approaches with radio interferometers; and (c) observation of the global CD/EoR signal using single antennas. Challenges in these detections involve removing foreground interference, achieving high-precision instrument calibration, and ensuring a high signal-to-noise ratio in both observation and data analysis for interferometers and single-antenna experiments.

The Square Kilometer Array (SKA, \citealt{koopmans2015cosmic}) is set to become the world's largest ground-based radio telescope, with CD/EoR detection as one of its key projects. SKA plans to conduct deep imaging of five specific sky fields and to statistically measure CD/EoR structures through extensive field surveys. SKA pathfinders, such as the Murchison Widefield Array (MWA, \citealt{barry2019improving}), Low Frequency Array (LOFAR, \citealt{mertens2020improved}), Giant Metrewave Radio Telescope (\citealt{paciga2013simulation}), and 21CMA (\citealt{zheng2016}), focus on statistical measurement of CD/EoR. However, due to sensitivity constraints, these pathfinders are currently unable to perform tomographic measurements of CD/EoR. Significant work in statistical CD/EoR measurements is still required (\citealt{shaw2023probing}; \citealt{2023arXiv231105364M}; \citealt{2023MNRAS.521.5120K}; \citealt{2020JAI.....950008D}; \citealt{2021MNRAS.500.2264H}), with numerous challenges remaining in data reduction and observational techniques.

Meanwhile, numerous experiments have been conducted to detect the global CD/EoR signal using individual antennas. Some of these experiments include EDGES (\citealt{bowman2008toward}), SCI-HI (\citealt{voytek2014probing}), BIGHORNS (\citealt{sokolowski2015bighorns}), LEDA (\citealt{bernardi2015foreground}), SARAS (\citealt{patra2013saras}), PRIZM (\citealt{philip2019probing}), and REACH (\citealt{de2022reach}).  In a study by \cite{bowman18}, a potential CD signal feature was reported by EDGES. This feature is characterized by an absorption feature at 78.2 MHz with an amplitude of approximately 556 mK. However, these findings, which challenge existing physical models, have recently been contested by results from the SARAS3 experiment
(\citealt{bevins2022astrophysical}). Further corroboration of these findings is needed through additional experimental observations.

In addition to the detection of the global signal from a single antenna, \cite{vedantham2015lunar} demonstrated a method for detecting the global redshifted 21-cm signal using interferometry, noting that short-spaced interferometers respond significantly to the global signal. \cite{presley2015measuring} discussed designing a global signal interferometer and performed simulations to evaluate its performance. \cite{2023ApJ...945..109Z} explored the possibility of extracting the 21 cm CD signal from measurements of the interferometer array, focusing on the feasibility of recovering the global spectrum from visibility data. \cite{singh2015detection} developed the theoretical framework for interferometer responses to global signals using dipole and resonant loop antennas, estimating their sensitivity. \cite{mckinley2020all} utilized an array of short-spacing antennas to detect the global CD signal interferometrically, asserting the theoretical feasibility of extracting the global redshifted 21-cm signal using this method. \cite{thekkeppattu2022system} reported on the Short Spacing Interferometer Telescope (SITARA), a prototype designed to measure global CD/EoR signals. Their further studies (\citealt{thekkeppattu2023singular}) delve into calibration, data analysis methods, and various factors influencing the performance of SITARA.

In this study, we present an experimental project, which is planned to be established in China, named the Short-spacing Interferometer Array for Global 21cm Signal Detection (SIGMA). This initiative is designed to detect the global HI 21cm signal using an interferometric approach with short-spacing antennas. A key aspect of our design involves addressing the challenge of mutual antenna coupling, a significant consideration in short-spacing antenna configurations as highlighted by \cite{mckinley2020all}. 

The structure of this paper is outlined as follows:
Section~\ref{sect:2} introduces the concept of interferometrically measuring the global CD/EoR signal. In Section~\ref{sect:3}, we provide detailed information on the design of the antennas and their layout. Site selection is also briefly introduced in this section, along with an overview of the electronic design of SIGMA. Section~\ref{sect:ct} delves into the effect of cross-talk. Finally, Section~\ref{sect:6} presents discussions and conclusions.

\section{Detecting the global signal with closely spacing interferometer}
\label{sect:2}
\begin{figure}
\centering
\hspace{-4mm}
\includegraphics[scale=0.8]{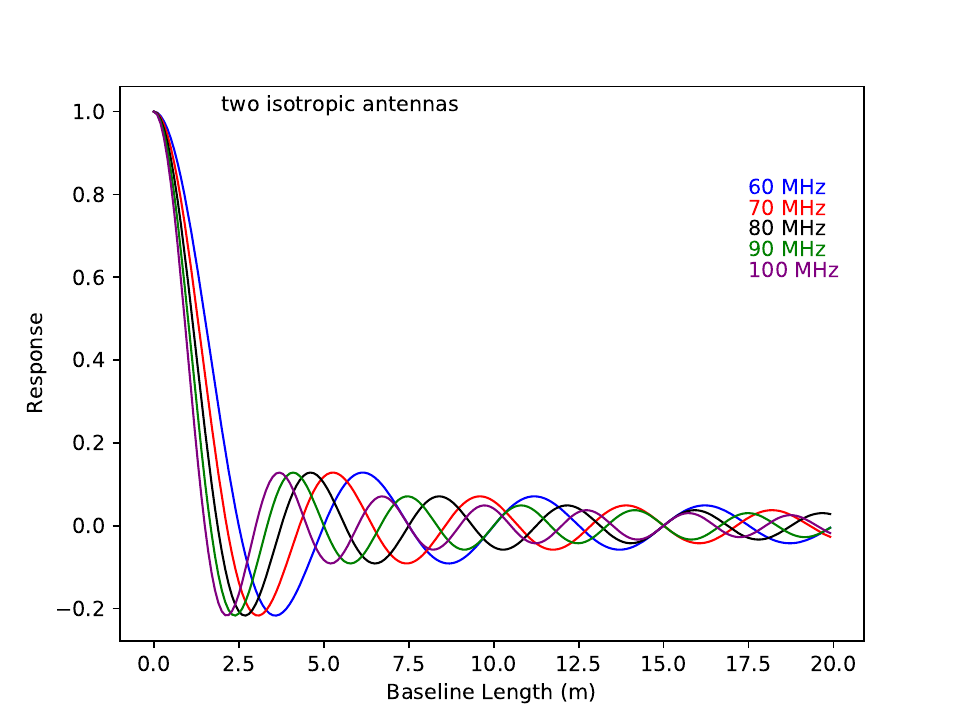}
\vspace{3mm}
\caption{Responses of two isotropic short spacing antennas at different frequencies. }
\vspace{3mm}
\label{fig:Fig1}
\end{figure}

Global 21cm signal can be measured with an interferometer when baselines are short enough to have a response to the global signal. The response of a two-element interferometer can be written as \cite{thompson2017interferometry}:
\begin{equation}
V(\boldsymbol{b},{\nu})=\frac{1}{4\pi}T_{\mathrm{sky}}({\nu})\int{A(\boldsymbol{r},{\nu})\mathrm{e}^{-2\pi i(\boldsymbol{b}\cdot \boldsymbol{r}/\lambda )}\mathrm{d}\Omega ,}
\label{equation1}
\end{equation}  
where V is the visibility in the unit of K. $b$ is the baseline vector and $\nu$ is the observation frequency. $T_{\rm sky}$ is the brightness temperature of the sky. $r$ is the unit vector of the sky direction. $A$ is the beam pattern of the antenna. $\lambda$ is the wavelength. $\Omega$ is the solid angle. Figure~\ref{fig:Fig1}, similar to Figure 1 in \cite{vedantham2015lunar}, shows that the interferometer's response to a global signal exhibits a \textit{sinc} shape as a function of baseline length across several different frequencies. When the spacing of the antenna elements is less than half the wavelength, there is a significant interferometric response to the global signal.

As described in \cite{mckinley2020all}, for real observation, the sky brightness temperature is not uniform but consists of a global component and angular variations. Consider $T_{\mathrm{angular}}(\boldsymbol{r},\nu )=T_{\mathrm{sky}}(\boldsymbol{r},\nu )-T_{\mathrm{sky}}(\nu )$ and the visibility equation can be written as:
\begin{equation}
\begin{aligned}
	V(\boldsymbol{b},{\nu})&=\frac{1}{4\pi}\bigl( T_{\mathrm{sky}}(\nu)\int{A(\boldsymbol{r},{\nu})\mathrm{e}^{-2\pi i(\boldsymbol{b}\cdot \boldsymbol{r}/\lambda})\mathrm{d}\Omega \quad}+\int{T_{\mathrm{angular}}(\boldsymbol{r},{\nu})A(\boldsymbol{r},{\nu})\mathrm{e}^{-2\pi i(\boldsymbol{b}\cdot \boldsymbol{r}/\lambda})\mathrm{d}\Omega} \bigr)\\
	&=T_{\mathrm{sky}}(\nu)\times \mathrm{global}\ \mathrm{response}+\mathrm{angular}\ \mathrm{response},\\
\end{aligned}
\label{equationspr}
\end{equation}
Short baselines respond better to large-scale structures and are insensitive to small angles, and we can assume that the first term of the equation dominates visibility. The $T_{\rm sky}(\nu)$ can then be estimated from the global response and visibility.


\section{Experiment Design}
\label{sect:3}

\subsection{Site selection}
\begin{figure}
\centering
\includegraphics[scale=0.35]{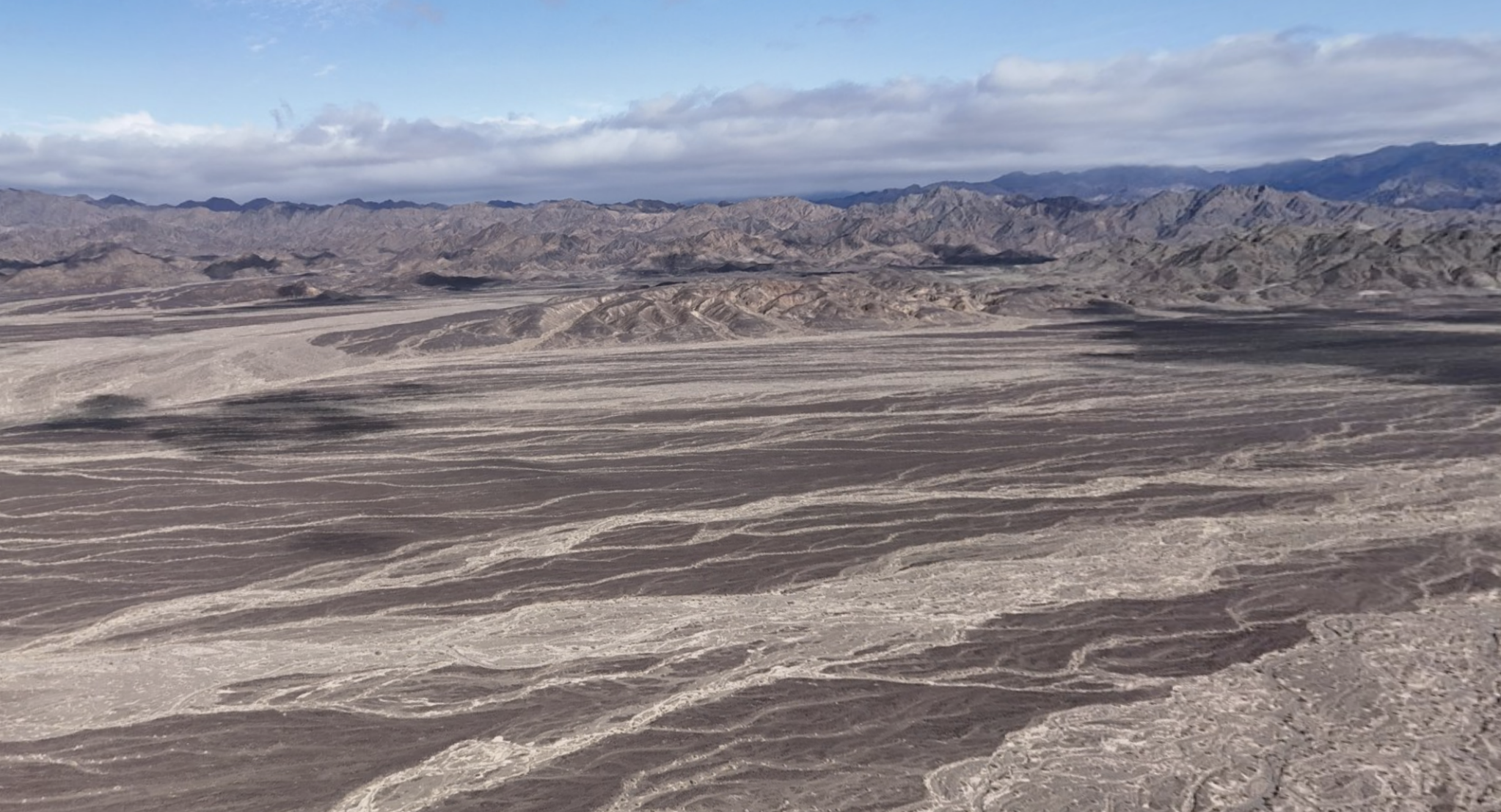}
\caption{Location of candidate sites near the city of Shanshan.}
\label{fig:site}
\end{figure}

\begin{figure}
\centering
\includegraphics[scale=0.8]{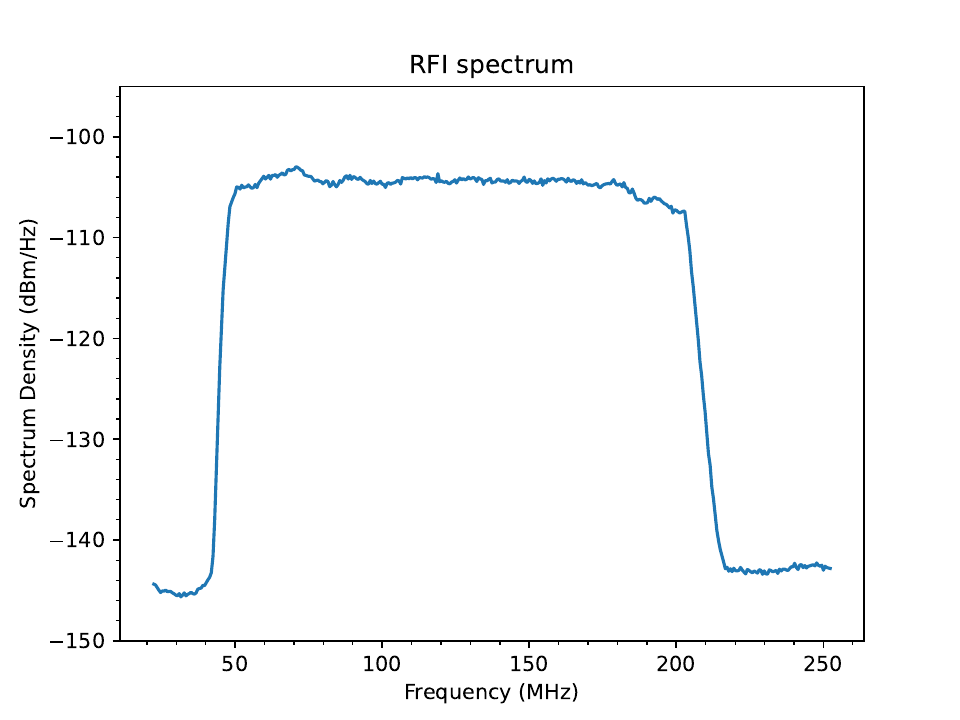}
\caption{The measured RFI spectrum was amplified by approximately 60dB with the first and second stages of amplifiers along band-pass filters covering the frequency range of 50-200 MHz used in the SIGMA system. }
\label{fig:rfi_sp}
\end{figure}

The experiment requires an RFI-quiet location that balances several considerations: ease of regular access for construction and resupply, and ease of power supply for the data processing center from a nearby power grid. Several potential sites have been surveyed in northwest China, including the Qinghai, Gansu, and Xinjiang provinces. One of the best candidates is near the city of Shanshan (${\rm E 90^\circ 12' 51.71'' ,\ N42^\circ 51' 38.84''}$) in Xinjiang province (Figure \ref{fig:site}). This site is about three hours from the nearest city. We measure the level of RFI around the site twice by surveying with a dipole antenna optimized for the frequency range $50\sim 200$ MHz,  two stages of amplifiers used in the SIGMA signal chain, and a \textit{Keysight} FieldFox RF analyzer N9913A, which means that the level of RFI was amplified by approximately 60 dB before being recorded by the spectrum analyzer. Figure~\ref{fig:rfi_sp} shows the radio spectrum between 50 and 200 MHz, amplified as mentioned, with bandpass filters covering the frequency range of 50-200 MHz. The spectrum has a bin size of 15 kHz, and the data is averaged over ten measurements using the average trace detector. The results suggest that the candidate site is an excellent and quiet location for low-frequency radio observation.

\subsection{Antenna array design}
\label{sect:3.1}

The redshifted 21cm signal from the CD/EoR is observable within the 50-200 MHz frequency range, corresponding to wavelengths between 6 meters and 1.5 meters. For our interferometer to effectively capture the global signal, the baselines must be shorter than half the wavelength, equating to less than 3 meters at 50 MHz. According to the findings of \cite{bowman18}, the absorption valley is located near 78 MHz. Therefore, SIGMA will focus on the 65-90 MHz band in the current phase.

\begin{figure*}
\centering
\includegraphics[scale=0.7]{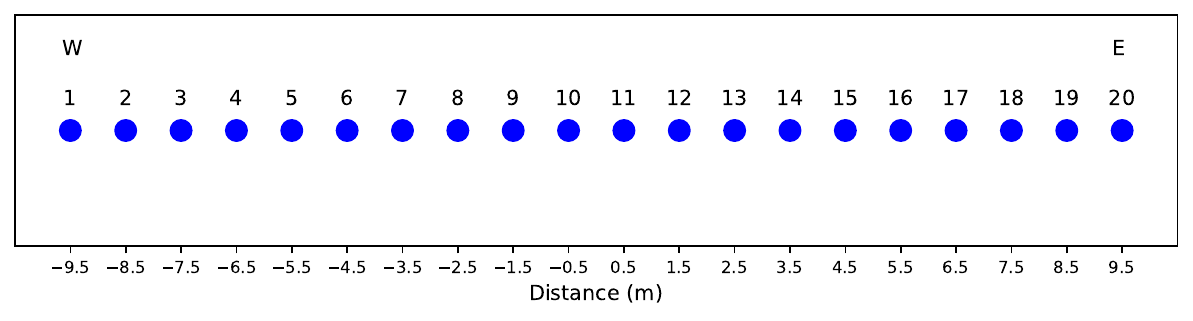}
\caption{The figure shows the layout of the SIGMA array. The array is arranged east-west, each blue dot corresponds to an antenna, the antennas are 1 meter apart, and the numbers on the blue dots are the numbers of the antennas.}
\label{layout1}
\end{figure*}

\cite{singh2015detection} proposed that a parallel configuration would yield enhanced sensitivity in antenna design. Here we also consider a parallel configuration 1-D array distributed along the E-W direction, as depicted in Figure~\ref{layout1}. This configuration consists of 20 antennas aligned in a straight line, oriented E-W, with a uniform distance of 1 meter between each antenna.

Here, we choose a one-dimensional (1-D) array for the following reasons. First, placing all antennas along a single baseline helps ensure each element is in a similar environment, which simplifies both the analytical modeling and the calibration of mutual coupling (Section~\ref{sect:ct}). Second, a linear array can be easily expanded or reduced, making it convenient for iterative experiments and allowing us to adjust the number of elements as needed. Third, since our goal is to measure the large-scale (global) signal rather than produce high-resolution images, we do not require the dense \emph{uv} coverage that often necessitates two-dimensional layouts. Consequently, a 1-D arrangement with short baselines meets our needs without introducing the more complex mutual coupling that can arise in multi-directional arrays.

In this work, we present a reference design with 20 antennas spaced 1 m apart along an East-West line (Figure~\ref{layout1}). This number is chosen mainly to illustrate our design principles and to evaluate antenna performance. However, it is not fixed; future experiments may adjust the array size depending on site conditions, resource availability, or sensitivity requirements.

\begin{figure}
\centering
\includegraphics[scale=0.8]{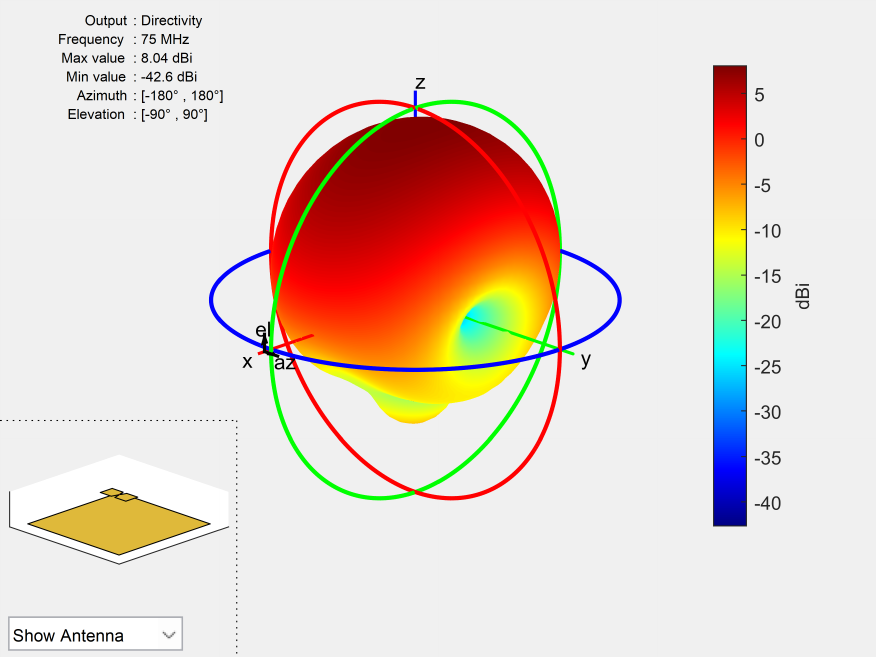}
\caption{Single blade antenna radiation pattern in 75 MHz.}
\label{beam1}
\end{figure}

\begin{figure}
\centering
\includegraphics[scale=0.7]{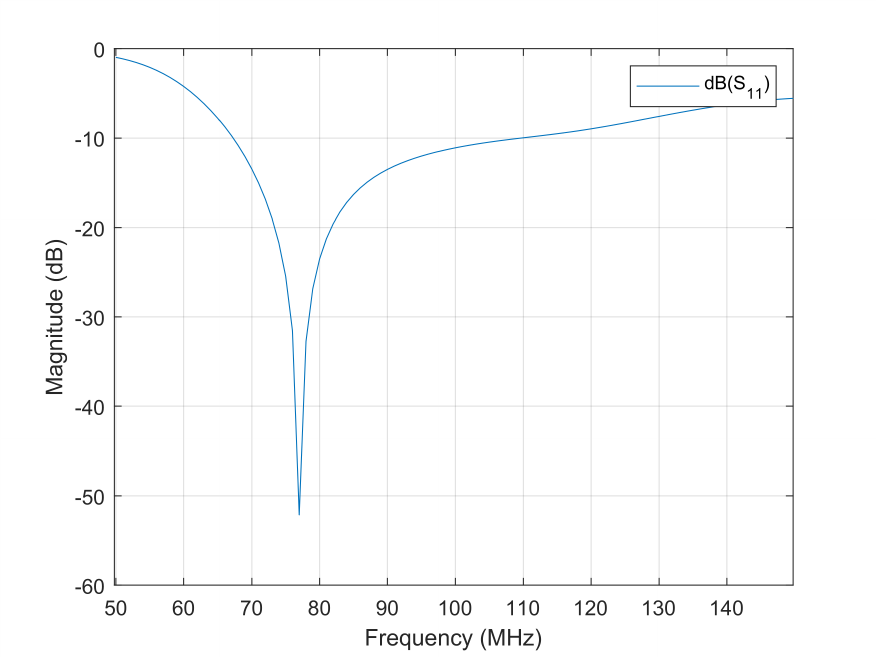}
\caption{Single blade antenna $S_{11}$.}
\label{S11single}
\end{figure}

In our systematic approach to designing antenna elements, we used a methodical process to identify appropriate antenna types, calculate relevant parameters, and assess various antenna designs using simulation. The simulated performance of the antenna will be one of the important criteria for our selection. Given our specific requirements and the power of simulation, we initially selected and evaluated the blade antenna.
 The blade antenna is favoured for its simple construction, wide bandwidth, ease of manufacturing, and near-omnidirectional radiation pattern, which makes it particularly advantageous for global signal detection. A typical blade antenna element in our design consists of two symmetrical blade panels, each 0.7 meters long, 0.65 meters wide, and the taper length is 0.6 meters. These elements are positioned 1.2 meters above the ground, and a 10-meter-wide, 30-meter-long metal mesh was placed on the ground. The parameter $S_{11}$, also known as input return loss, reflects the amount of power that is reflected back from the antenna due to impedance mismatch. A lower value $S_{11}$ is preferable, as it indicates that a greater proportion of the power is effectively transmitted into the antenna, which is crucial for signal capture in our experiment. The radiation pattern and the $S_{11}$ of the isolated single blade antenna are shown in Figure~\ref{beam1} and Figure~\ref{S11single}. From the simulation results, it can be seen that the blade antenna performs well in the 65 to 90 MHz frequency band and is suitable for the SIGMA experiment.

\begin{figure*}
\centering
\includegraphics[scale=0.45]{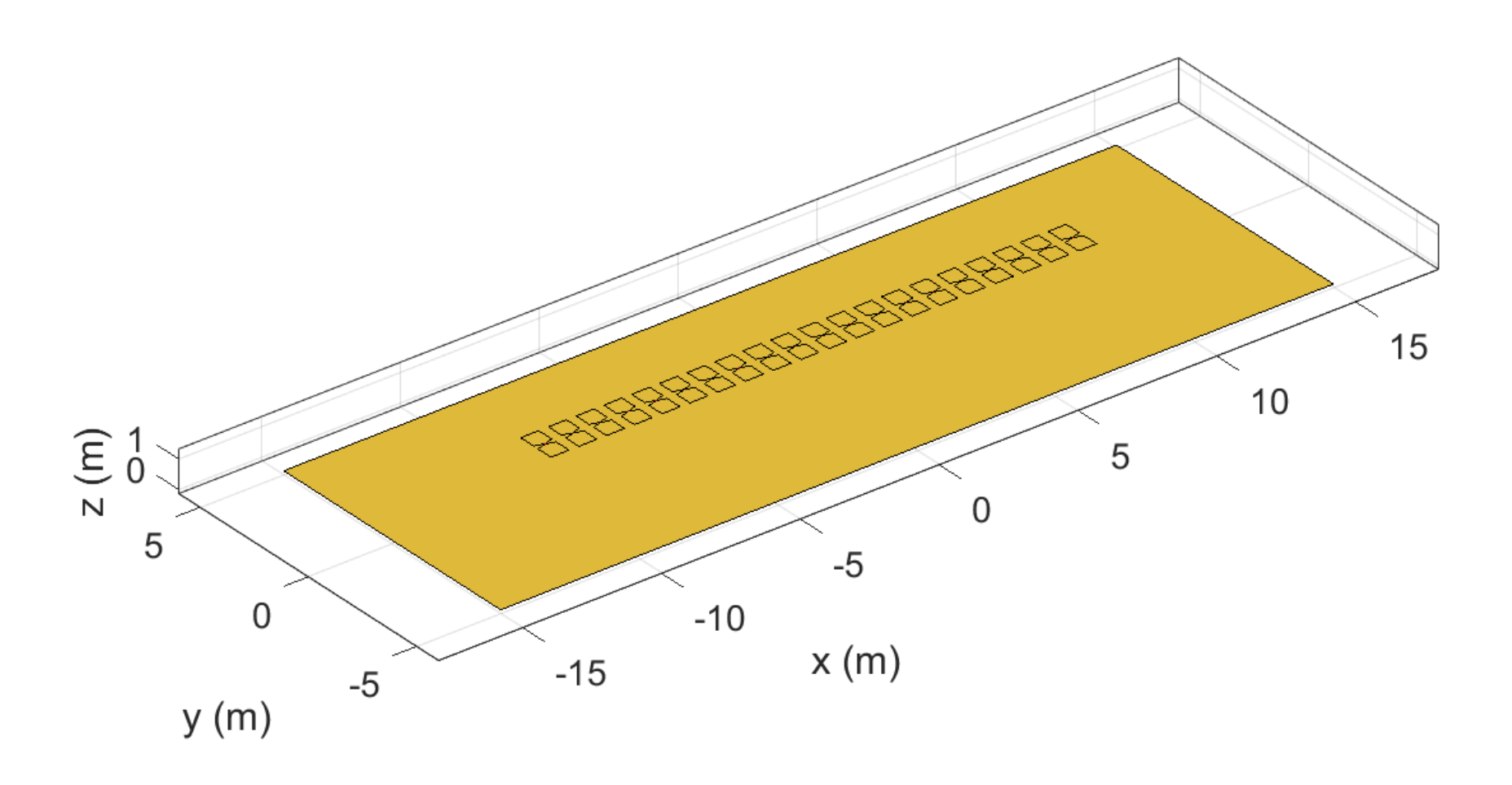}
\caption{3D view of the SIGMA 20 blade antennas model. The antennas are in parallel configuration with a height of 1.2 meters and the ground plane is $30 \times 10$ meters of PEC (Perfect Electric Conductor) material.}
\label{layout2}
\end{figure*}

We have modelled the antenna array using the \texttt{Matlab Antenna Toolbox}\footnote{\href{https://ww2.mathworks.cn/products/antenna.html}{https://ww2.mathworks.cn/products/antenna.html}}, the model is shown in Figure~\ref{layout2}. The information about the antenna and array design is summarized in Table~\ref{tab:SIGMA-params}.

\begin{figure*}
  \centering
  \includegraphics[scale=0.3]{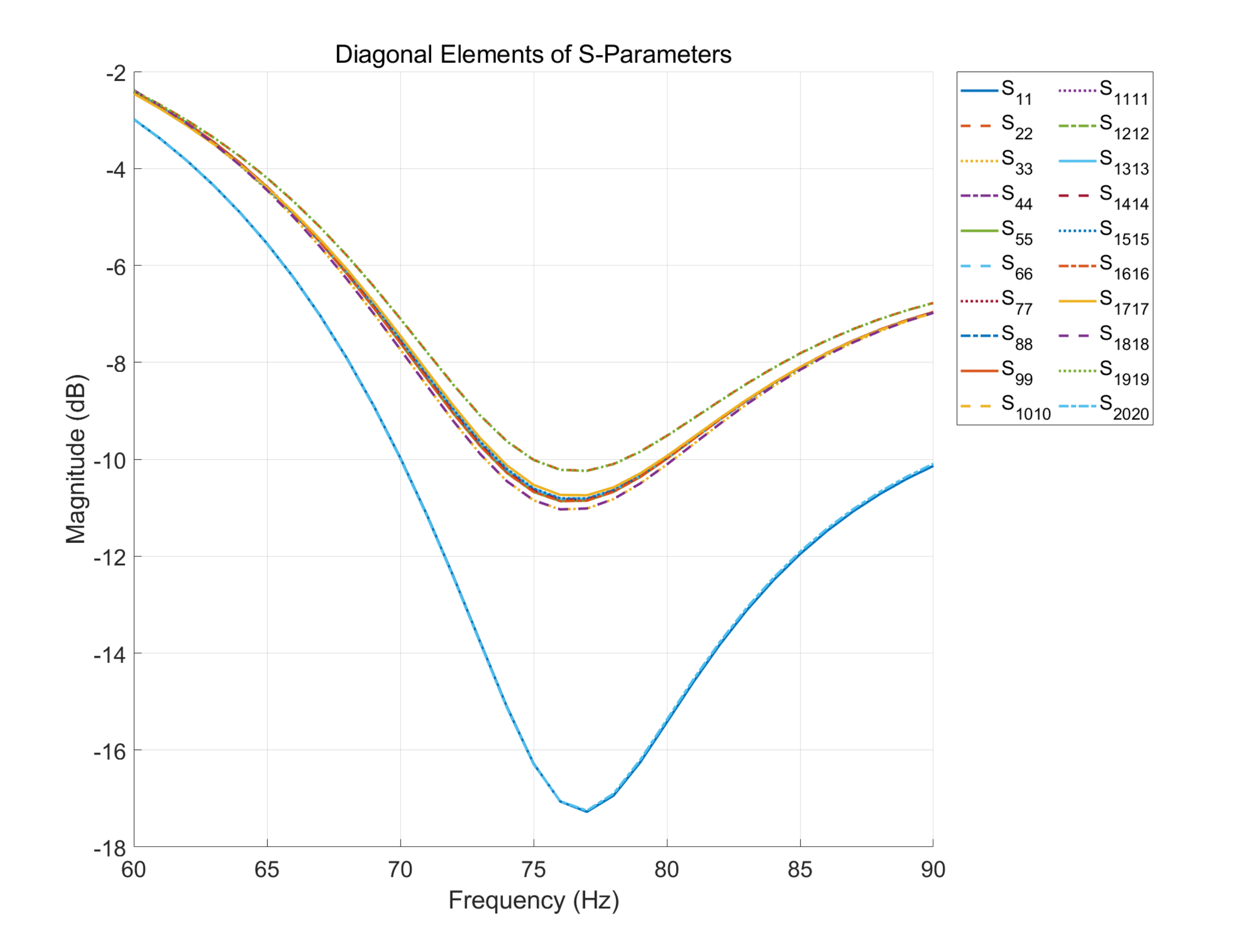}
  \caption{Simulation results of $S_{\rm nn}$ of the SIGMA. Due to the effect of mutual coupling, the $S_{\rm nn}$ are considerably transformed compared to the single antenna case, but they are still usable in 65-90 MHz and especially around 78 MHz. In this figure, $S_{11}$ and $S_{2020}$ overlap (the dark blue line at the bottom and the light blue dash-dot line), $S_{22}$ and $S_{1919}$ overlap (the orange-red dashed line at the top and the green dotted line),  $S_{33}$ and $S_{1818}$ overlap (the second line from the bottom, shown by the yellow dotted line and the purple dashed line). The remaining traces from $S_{44}$ through $S_{1717}$ essentially overlap one another, indicating that antennas 4 through 17 exhibit nearly identical $S_{\rm nn}$.}
  \label{SNN}
  \end{figure*}

Here we simulated the $S_{\rm nn}$ (input return loss of antenna n) of each antenna of the SIGMA, the termination resistance is internally set to a default of 50 ohms, and the results $S_{\rm nn}$ are shown in Figure~\ref{SNN}. $S_{\rm nn}$ is essentially the same except for antenna 1, 20; 2, 19; 3, 18. Compared to the single antenna case, the $S_{\rm nn}$ change significantly due to the effect of mutual coupling, but the antenna is still usable in the 65-90 MHz band.

\begin{figure*}
     \centering
     \begin{subfigure}[b]{0.49\linewidth}
         \centering
         \includegraphics[width=\linewidth]{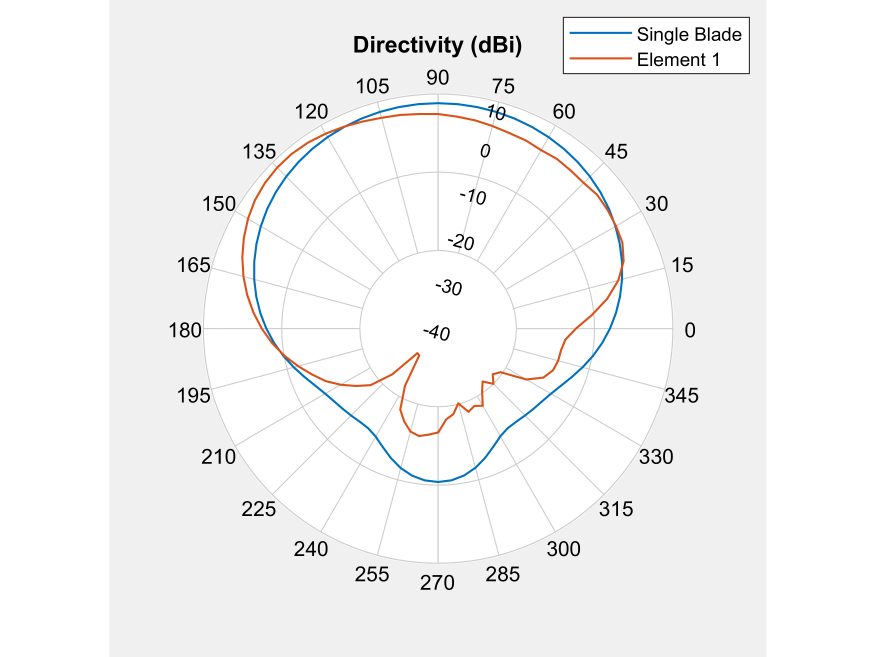}
         \caption{Element 1}
     \end{subfigure}
     \hfill
     \begin{subfigure}[b]{0.49\linewidth}
         \centering
         \includegraphics[width=\linewidth]{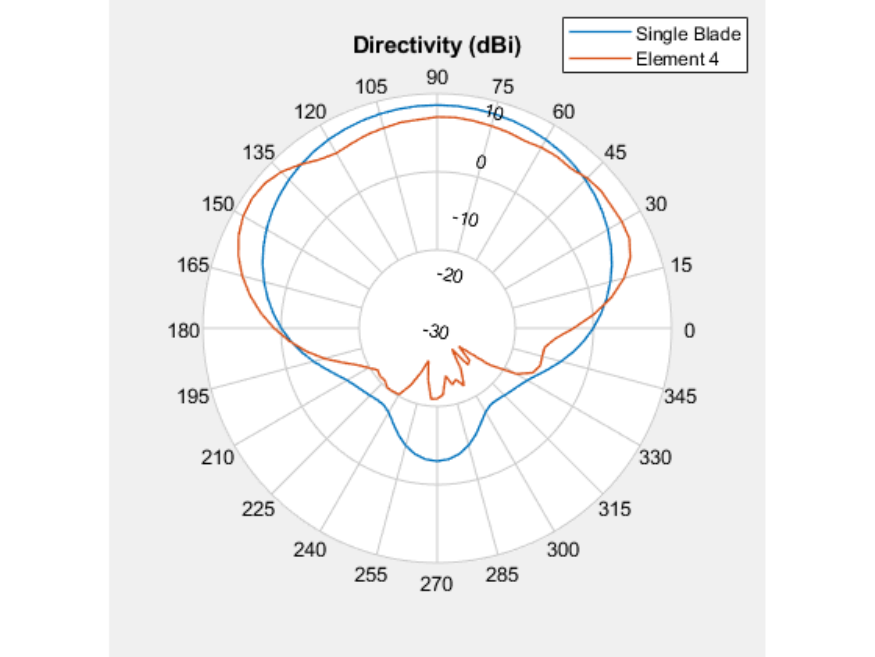}
         \caption{Element 4}
     \end{subfigure}
     
     \begin{subfigure}[b]{0.49\linewidth}
         \centering
         \includegraphics[width=\linewidth]{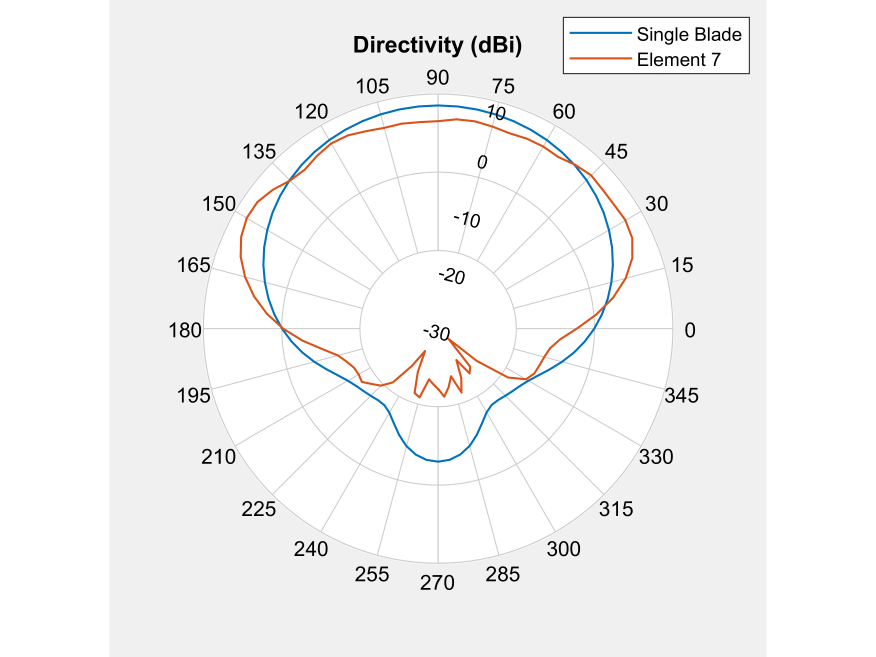}
         \caption{Element 7}
     \end{subfigure}
     \hfill
     \begin{subfigure}[b]{0.49\linewidth}
         \centering
         \includegraphics[width=\linewidth]{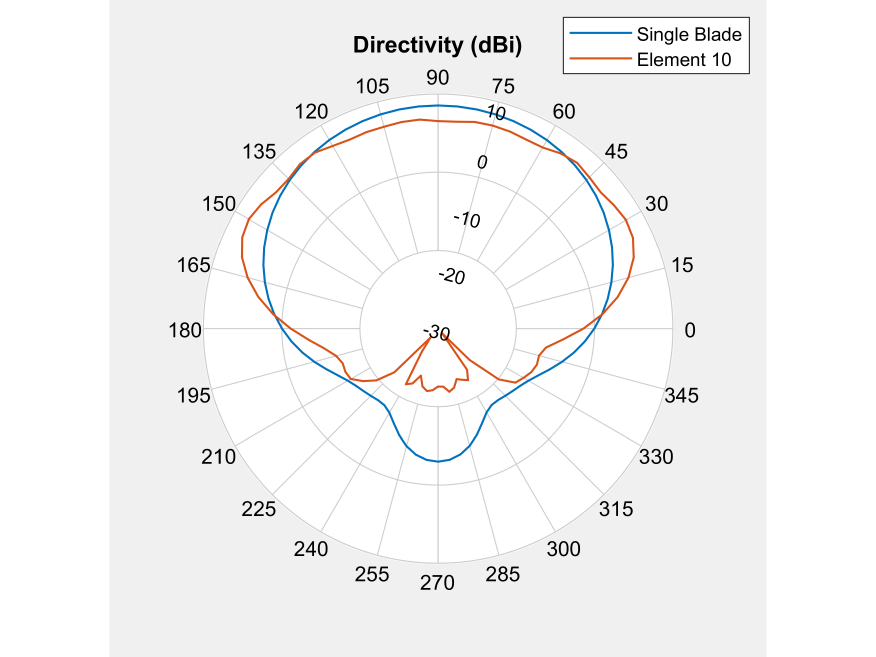}
         \caption{Element 10}
     \end{subfigure}
     \caption{Comparison of the radiation pattern of elements 1, 4, 7, and 10 in the array (EW direction) with single blade antenna in dBi, the top corresponding to the zenith. Antennas in the array still respond well to a wide range. There are some variations in the radiation patterns of the different antennas due to the antenna positions, especially for antenna 1 (symmetrical, antenna 20). In our experiment, radiation patterns below the plane of the antenna ($180^\circ - 360^\circ$) do not affect observations.}
     \label{FigBEAM}
\end{figure*}

Furthermore, the antenna radiation pattern is a critical element that warrants detailed consideration. We simulated the radiation pattern of antenna numbers 1, 4, 7 and 10, the results are shown in Figure~\ref{FigBEAM}. The radiation pattern is calculated by driving a specific element in the array. The rest of the array elements are terminated using reference impedance (50 ohms).

\begin{table}[!htb]
\centering
\caption{Key parameters of the SIGMA design}
\label{tab:SIGMA-params}
\begin{tabular}{lll}
\hline
\textbf{Parameter}       & \textbf{Value}         & \textbf{Notes}                           \\
\hline
Frequency range          & 65--90\,MHz            & Focus on global EoR signal               \\
Number of antennas       & 20                     & 1D linear array (E--W direction)         \\
Antenna spacing          & 1\,m                   & Short baselines for detecting large-scale signal \\
Antenna type             & Blade antenna          & Near-omnidirectional, wide bandwidth     \\
Blade size         & 0.7\,m (L), 0.65\,m (W), 0.6\,m taper & Two symmetric blade panels \\
Antenna height           & 1.2\,m                 & Above ground                             \\
Ground plane             & 10\,m\,$\times$\,30\,m & Metal mesh (PEC in simulation)           \\
\hline
\end{tabular}
\end{table}

In short-spacing interferometer experiments, the antenna spacing is required to be as small as possible in order to obtain as strong a response as possible, which also leads to strong mutual coupling effects. From the simulation results, it can be seen that the $S_{\rm nn}$ and radiation patterns of the antennas change significantly at very close distances, especially for antenna 1, 20; 2, 19; 3, 18, whose performances are more different from those of the other antennas. An approach would be to use these antennas as constraints only, during the actual observation, their observation data will not be used.

The one-dimensional array design allows SIGMA to be easily expanded in the future and higher sensitivity can be obtained by continuing to install more antennas. The scale depends on the condition of the final selected telescope site. The effect of mutual coupling, which will be stronger for shorter baselines, needs to be properly dealt with to measure the global EoR signals. We try to develop a framework in Section \ref{sect:ct} and hope to remove it in the calibration.  The similar design will be implemented for the high band ($90\sim 200$ MHz) in the next phase of SIGMA.

\subsection{Electronics design}
\begin{figure}
  \centering
  \includegraphics[scale=0.6]{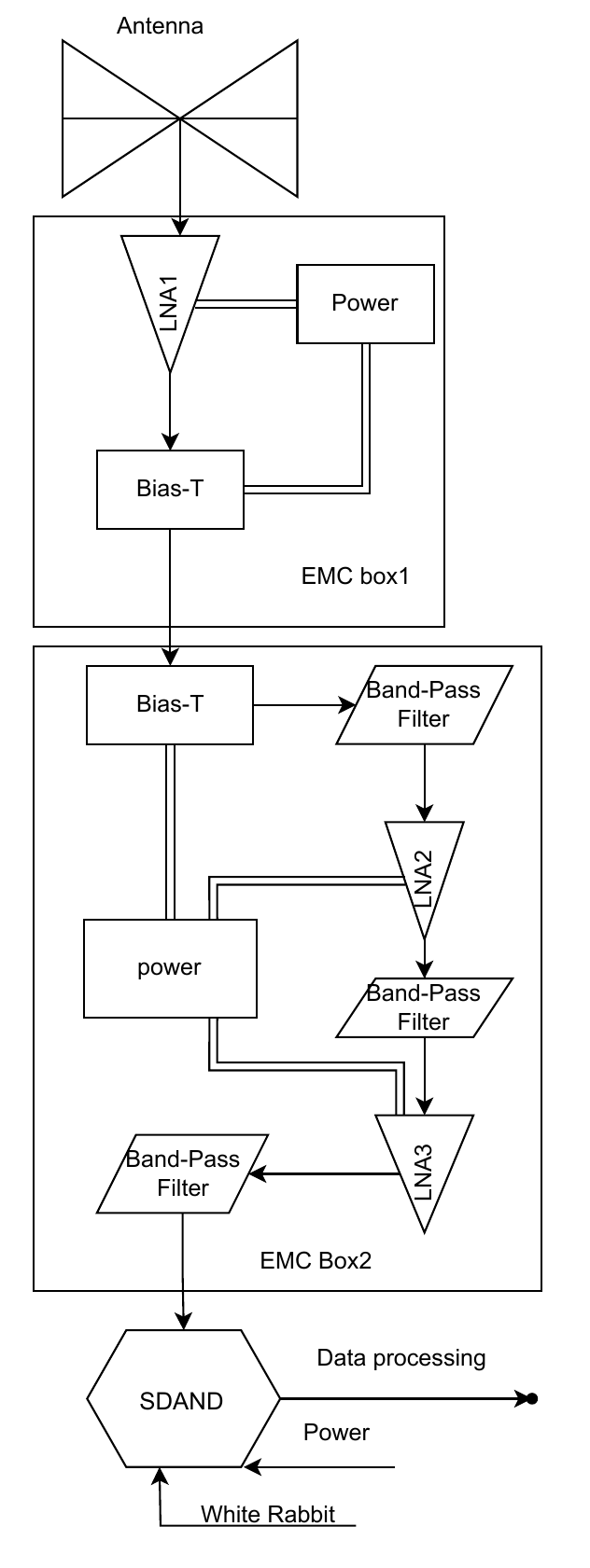}
  \caption{The block diagram showing the electronics system for one antenna in the SIGMA system.}
  \label{fig:block}
  \end{figure}

In summary, the SIGMA system consists of antennas kept in close proximity, a local 'fieldbox' performing analog signal conditioning, digitization equipped with a time-synchronization client from the White Rabbit system and a remote correlation centre with GPUs performing the receiving and correlation for all antennas.No hardware calibration device such as noise diodes is integrated yet. The initial prototype is designed to be as simple as possible, focusing on studying systematics such as cross-talk between antennas. A block diagram of the SIGMA electronics system is given in Figure \ref{fig:block}, and the block diagram of SIGMA system and data processing is shown in Figure \ref{sys}.

In the initial prototype system, each antenna is associated with a local fieldbox via a coaxial cable. In the fieldbox, the first stage low noise amplifier(LNA) is assembled based on the chip XT3074 from Chengdu Fairchild Technology Ltd., with a gain of about 27 dB and a noise figure of about 0.5 dB. Through the same coaxial cables the DC power for the first stage LNA is supplied by the ﬁeldbox via bias-tees, which makes the deployment of the fieldbox flexible in terms of location. The second and third stage amplifiers are built based on BFP740ESD from INFINEOON, with a gain of about 40 dB and a noise figure of less than 1.2 dB, and ZD3027 from CEmaxRF tech, with a gain of about 25 dB and a noise figure of less than 1.6 dB, respectively. 
To reduce the effects of out-of-band radio frequency interference (RFI) on signal chain linearity, since the amplifiers used are broadband relative to the required 200 MHz bandwidth, a customized 50-200 MHz band-pass filter with suppression greater than 50 dB outside the band-pass is applied between the amplifier stages and before analog-to-digital conversion (ADC). Together, the analog section has a net gain greater than 80 dB and is shielded by a metal EMC box to reduce interference from outside sources.

The amplified and filtered signals are then digitized in data acquisition nodes, utilizing a customized Synchronized Distributed Array Node Digitizer (SDAND) system. The SDAND system was designed by our team, and further details of the SDAND system will be presented separately (Gu et al., in preparation). This system comprises a set of identical data acquisition units, whose sampling clocks are synchronized using the White Rabbit (WR) system \citep{lipinski2011white} with sub-nanosecond accuracy. The SDAND system features several input/output ports, including two coaxial connectors for RF signal input, one optical fiber connector for transporting WR system signals, one optical fiber connector for receiving control messages, and two optical fiber connectors for outputting payload data. The advantages of the SDAND system are threefold: 1) it enables the formation of a versatile range of radio telescopes, 2) it facilitates the creation of an extensive and flexible data acquisition system with stability, and 3) it transparently handles time synchronization among all data acquisition nodes. Since nearly all types of array-formed radio telescopes rely on synchronously sampling voltage signal outputs from the antenna system, the output data stream of the SDAND system can be appropriately directed to realize various array telescopes, including single beam-forming stations, beam-forming based interferometers (e.g., the SKA LFAA), wide-field monitoring interferometers (e.g., LWA and HERA), pulse-signal triggering and event reconstructing detectors for particle physics studies (e.g., GRAND, \citealt{martineau2017giant}), and more. As the signals are transferred in digital mode, data processing centres equipped with high-performance processors and GPUs can be situated far from radio-quiet zones, ensuring sufficient resources for computing and cooling. With the SDAND system's scalability and flexibility, expanding or repurposing the SIGMA system is relatively straightforward. This can be achieved by adding more antennas with SDAND devices, modifying the software, and increasing the GPU cluster's data acquisition and correlation computing capabilities as needed.

The prototype of the SIGMA system comprises 7 data acquisition nodes,  each receiving signals from two RF signal processing chains, along with a data processing node equipped with four Nvidia A40 GPU cards and two  100 Gbps Ethernet interface cards. All the data acquisition nodes and the data processing node are connected to a 100 Gbps switch, with the processing node and Ethernet switch located far away from the radio quiet zone. The data acquisition nodes stream the sampled payload data to the processing node. Real-time channelization and correlation are performed on GPU cards of the data processing node to generate auto- and cross-correlation spectral products. The acquired time domain data series are channelized into 65,536 spectral channels. Since the data acquisition system is a direct sampling system, each data point is a real number, resulting in only 32,768 independent channels spanning a frequency range of 0–200 MHz, with a spectral resolution of about 6.1 kHz. With the bandpass filter in the RF signal chain, only signals of frequency between 50 and 200 MHz are sensitive to sky signals. All the data acquisition nodes and the data processing node are connected to a  100 Gbps Ethernet switch. The payload data streams from the eight acquisition nodes (i.e., 14 data streams) are routed to the data processing node, where the auto- and cross-correlation products are computed. The computing tasks are evenly distributed among the four GPU cards, with each GPU processing data acquired within a quarter of the time. The correlation results from the four GPU cards are gathered and summed to form the final results of each time step, which are then stored on disks. The information about the electronic design is summarized in Table~\ref{tab:electronics}.

\begin{table}[htb]
\centering
\caption{Key electronics parameters and components in the SIGMA prototype}
\label{tab:electronics}
\begin{tabular}{p{3.7cm}p{4.0cm}p{6.1cm}}
\hline
\textbf{Component / Stage} & \textbf{Model / Spec} & \textbf{Notes} \\
\hline
\textbf{First-stage LNA} 
& XT3074 (Chengdu Fairchild Tech) 
& Gain $\sim27$\,dB, Noise Figure (NF) $\sim0.5$\,dB \\

\textbf{Second-stage amp} 
& BFP740ESD (Infineon) 
& Gain $\sim40$\,dB, NF $<1.2$\,dB \\

\textbf{Third-stage amp} 
& ZD3027 (CEmaxRF Tech) 
& Gain $\sim25$\,dB, NF $<1.6$\,dB \\

\textbf{Band-pass filter} 
& 50--200\,MHz 
& Suppression $>50$\,dB outside 50--200\,MHz \\

\textbf{Number of SDAND nodes}
& 7 
& Each SDAND handles 2\,RF signal chains, total 14\,RF inputs \\

\textbf{Sampling scheme} 
& 
& 0--200\,MHz sampled, yields 65,536 channels (32,768 independent), $\sim$6.1\,kHz spectral resolution \\

\textbf{Correlation center} 
& GPU-based server 
& 4$\times$\,NVIDIA A40 GPU cards, 2$\times$\,100\,GbE NICs \\

\textbf{Network switch} 
& 100\,GbE switch 
& Aggregates 7 SDAND data streams (14 total) for correlation \\

\hline
\end{tabular}
\end{table}

\section{Cross-Talk}
\label{sect:ct}

  \begin{figure*}
  \centering
  \includegraphics[scale=0.48]{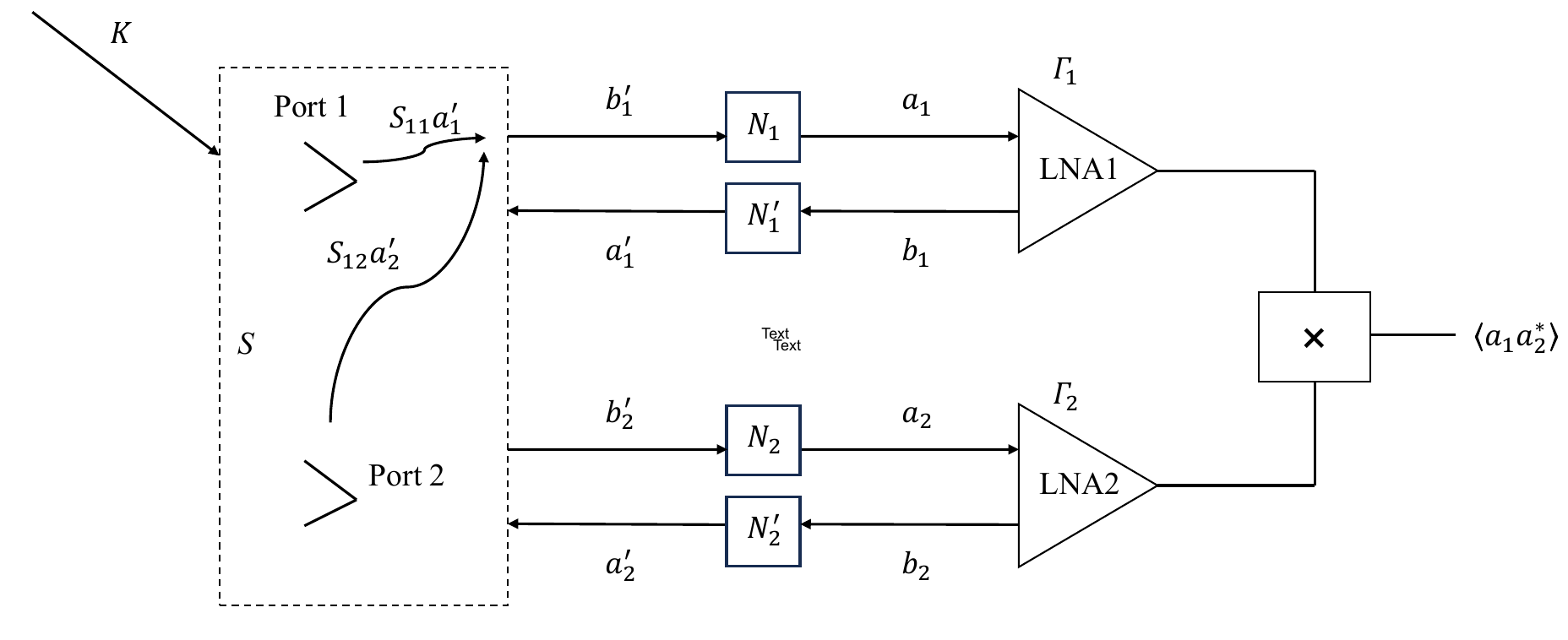}
  \caption{Simplified two-antenna interferometer model, where the signal reflects at the LNA, the reflected signal reflects again at the antenna port, and then leaks into the rest of the antenna system. For clarity, only the interference received by LNA1 is shown.}
  \label{model}
  \end{figure*}

\begin{figure*}
    \centering
    \begin{subfigure}[b]{0.49\linewidth}
        \centering
        \includegraphics[width=\linewidth]{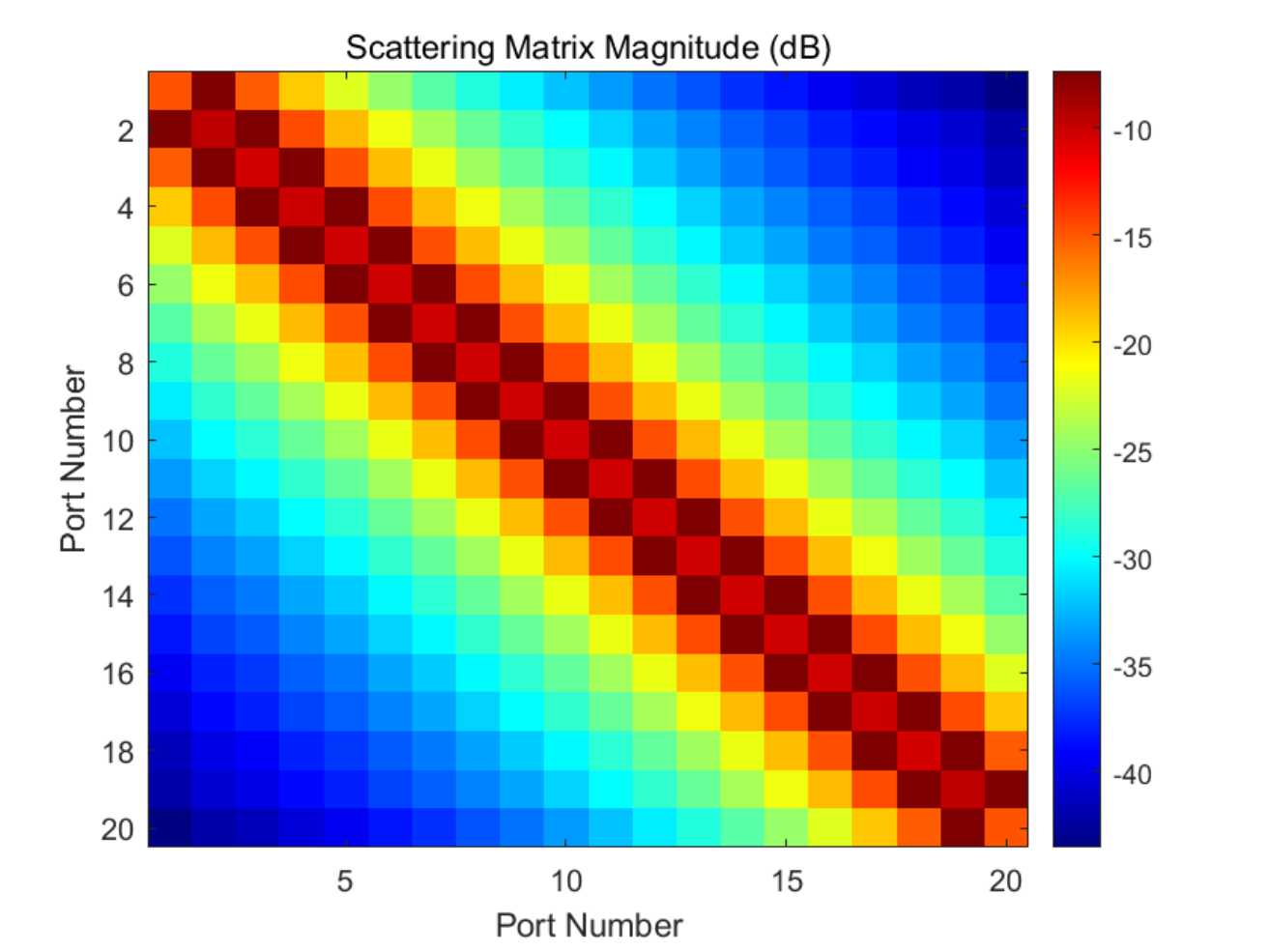}
        \caption{Magnitude plot of simulated SIGMA S parameters matrix in 75 MHz.}
        \label{fig:sm_magnitude}
    \end{subfigure}
    \hfill
    \begin{subfigure}[b]{0.49\linewidth}
        \centering
        \includegraphics[width=\linewidth]{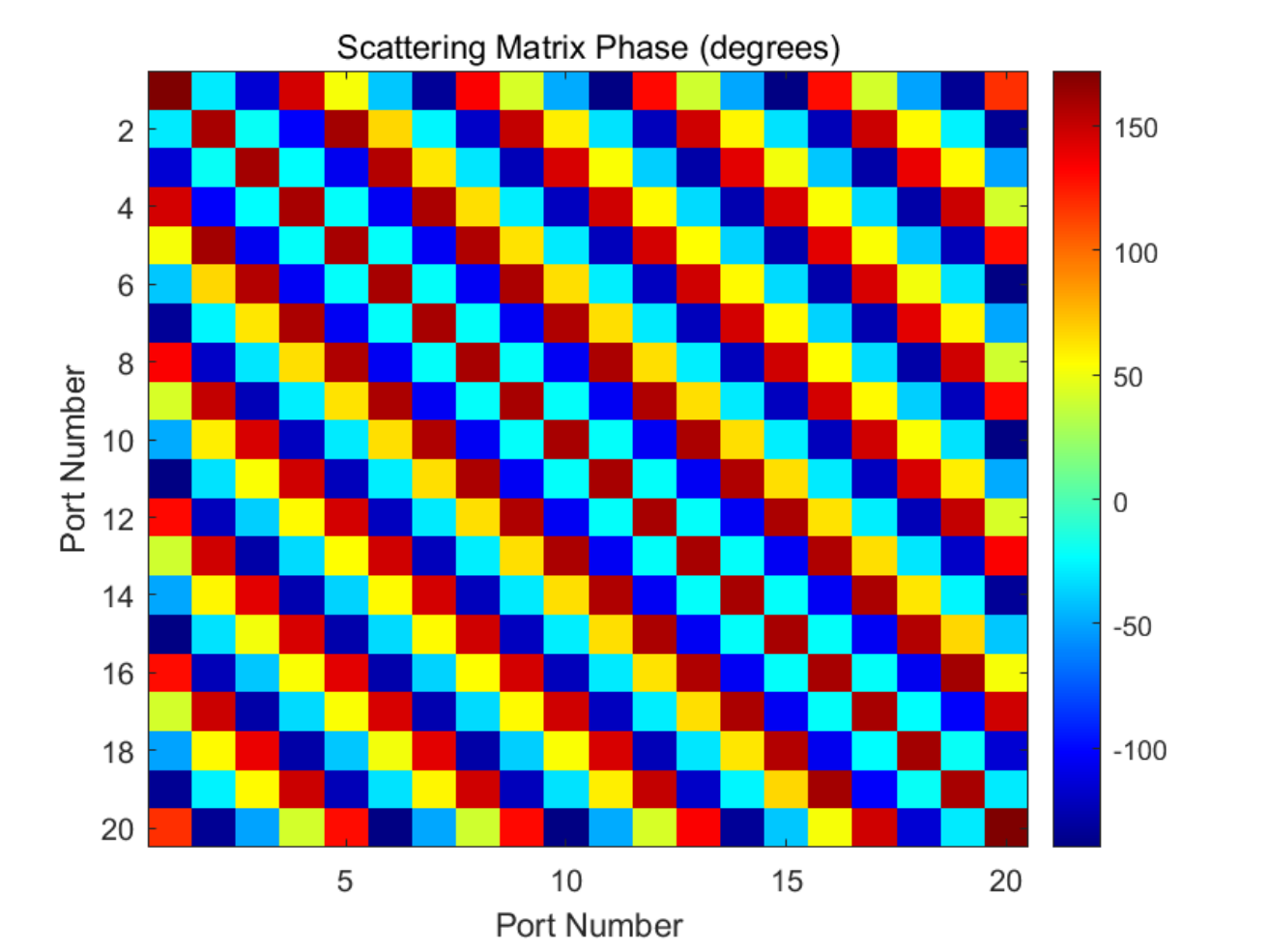}
        \caption{Phase plot of simulated SIGMA S parameters matrix in 75 MHz.}
        \label{fig:sm_phase}
    \end{subfigure}
    \caption{Left panel: Magnitude plot of simulated SIGMA S parameters matrix in 75 MHz. Right panel: Phase plot of simulated SIGMA S parameters matrix in 75 MHz.}
    \label{fig:sm_combined}
\end{figure*}

In our short-spacing interferometer array, the distance between antennas is less than half a wavelength, meaning that the interference between antennas cannot be neglected. \cite{thekkeppattu2022system} discusses how receiver noise may leak into adjacent antennas, resulting in non-zero coherent receiver noise. \cite{2023arXiv231103494K} developed a model for receiver noise due to cross-coupling between different antennas. In addition, we focus on the cross-talk from sky signals and propose a strategy to remove it.

Considering there are $n$ antennas, formed a network which has $n$ ports. Each antenna port is connected to a low noise amplifier (LNA). Assuming all interferences are negligible, let each port's reflected wave be denoted as $k_{\rm i}$, and let $K = (k_1, k_2, \cdots, k_{\rm n})^T$. So $K$ refers to the sky signal being reflected by the antenna.
The entire coherence matrix without cross-talk effect can be expressed as:

    \begin{align}
        \text{Coherence Matrix} = C = 
        \begin{bmatrix}
            \langle k_1 k_1^* \rangle & \langle k_1 k_2^* \rangle & \cdots & \langle k_1 k_{\rm n}^* \rangle \\
            \langle k_2 k_1^* \rangle & \langle k_2 k_2^* \rangle & \cdots & \langle k_2 k_{\rm n}^* \rangle \\
            \vdots & \vdots & \ddots & \vdots \\
            \langle k_{\rm n} k_1^* \rangle & \langle k_{\rm n} k_2^* \rangle & \cdots & \langle k_{\rm n} k_{\rm n}^* \rangle
        \end{bmatrix}
    \end{align}

Here $k_{\rm i}^*$ denotes the complex conjugate of $k_{\rm i}$.
And $C$ can also be written as:

    \begin{align}
        C = \langle (k_1, k_2, \cdots, k_{\rm n})^T (k_1^*, k_2^*, \cdots, a_{\rm n}^*) \rangle = \langle K K^* \rangle,
    \end{align}  
where $K^*$ denotes the conjugate transpose of $K$.
In our experiment, we are only interested in baselines shorter than half a wavelength, particularly the nearest baselines. We are specifically interested in the elements on the super-diagonal and its next two diagonals. 

In practical scenarios, due to the interference from the two effects previously mentioned, the signal received by the antenna 
is not simply $K$, further analysis is required. For each $\text{LNA}_{\rm i}$, denote its forward wave as $a_{\rm i}$ and its reflected wave as $b_{\rm i}$.
Each $\text{LNA}_{\rm i}$ also has its noise source, which can be represented as an equivalent forward wave $N_{\rm i}$ and a backward wave $N'_{\rm i}$ at the $\text{LNA}_{\rm i}$ input. Also, each port has a forward wave $a'_{\rm i}$ and a reflected wave $b'_{\rm i}$. Suppose $S = (S_{\rm ij})$ is the scattering matrix, where $S_{\rm ij}$ is the scattering coefficient from antenna $j$ to antenna $i$, implying that $S_{\rm ij}*a'_{\rm j}$ represents the scattering from antenna $j$ to antenna $i$.

Focusing on the two antenna ports as illustrated in Figure \ref{model}, for antenna port 1, its incident wave $a'_1$ includes the reflected wave $b_1$ of $\text{LNA}_1$ as well as its noise's equivalent backward wave $N'_{\rm i}$. Its reflected wave $b'_1$ includes the original sky signal $k_1$ along with interference generated from antenna coupling. For $\text{LNA}_1$, its incident wave $a_1$ includes $b'_1$ and its own noise's equivalent forward wave $N_1$. It records the signal and reflects a portion of it. Let $\varGamma_{\rm i}$ be the reflection coefficient of $\text{LNA}_{\rm i}$, then the reflected signal is $b_1 = \varGamma_1a_1$. Therefore, we have the following equations:
    \begin{align}
        \left\{
        \begin{array}{l}
            a_1=b'_1+N_1\\
            b_1=\varGamma_1a_1\\
            a'_1=b_1+N'_1\\
            b'_1=k_1 + S_{11}a'_1 + S_{12}a'_2 + \cdots + S_{\rm 1n}a'_{\rm n}  = k_1 + \sum{S_{\rm 1i}a'_{\rm i}}\\
        \end{array}
        \right
        .
    \end{align}
We denote
    \begin{align*}
        A &= (a_1, a_2, \ldots, a_{\rm n})^T, \\
        B &= (b_1, b_2, \ldots, b_{\rm n})^T, \\
        A' &= (a'_1, a'_2, \ldots, a'_{\rm n})^T, \\
        B' &= (b'_1, b'_2, \ldots, b'_{\rm n})^T, \\
        N &= (N_1, N_2, \ldots, N_{\rm n})^T, \\
        N' &= (N'_1, N'_2, \ldots, N'_{\rm n})^T.
    \end{align*}
Examining each port and LNA similarly, we can obtain
        \begin{align}
        \left\{
        \begin{array}{l}
            A=B'+N\\
            B=\varGamma A\\
            A'=B+N'\\
            B'=K + SA'\\
        \end{array}
        \right.
        ,\label{solveall}
    \end{align}
    where
    \begin{align}  
    S &=              
    \begin{pmatrix}
        S_{11} & S_{12} & \cdots & S_{\rm 1n} \\
        S_{21} & S_{22} & \cdots & S_{\rm 2n} \\
        \vdots & \vdots & \ddots & \vdots \\
        S_{\rm n1} & S_{\rm n2} & \cdots & S_{\rm nn}
        \end{pmatrix},
    \end{align}
    \begin{align}
        \varGamma &= 
        \begin{pmatrix}
        \varGamma_{1} & 0 & \cdots & 0 \\
        0 & \varGamma_{2} & \cdots & 0 \\
        \vdots & \vdots & \ddots & \vdots \\
        0 & 0 & \cdots & \varGamma_{\rm n}
        \end{pmatrix},
    \end{align}
    From (\ref{solveall}) we can obtain 
    \begin{align*}
        A = K + SA' + N = K + SB + SN' + N = K + S \varGamma A + SN' + N,
    \end{align*}
    thereby
    \begin{align}
        (I - S \varGamma )A = K + SN' + N,
    \end{align}
    where $I$ is the identity matrix. The matrix $I - S \varGamma$ is determined by the properties of the array itself and, as can be seen below, is assumed to be invertible in the numerical sense. Let $R = (I - S \varGamma)^{-1}$. Let $Q = SN' + N$, then $Q$ represents the portion of the interference originating from the receiver noise. We have
    \begin{align}
        A =(I - S \varGamma )^{-1}( K + SN' + N) = R (K + Q).
    \end{align}

In practical interferometric array measurements, we are generally unable to determine $A$ directly. Instead, we can only measure the results of their coherence.

    \begin{align}
        \widetilde{C}
        &= \langle A A^* \rangle
        = R\langle (K+Q) (K^*+Q^*) \rangle R^*.\\
        &= R\langle KK^* \rangle R^* + R\langle  KQ^* + QK^* + QQ^* \rangle R^* \nonumber\\
        &= RCR^*  + R\langle  KQ^* + QK^* + QQ^* \rangle R^*.
        \label{CCc0}
    \end{align}

Our objective is to compute the undisturbed matrix $C$ from the obtained $\widetilde{C}$ through a specific method.
If we disregard the noise produced by the LNA, we can approximately obtain
    \begin{align}
        \widetilde{C} \approx RCR^*.
        \label{CCc}
    \end{align}

Evidently, to calculate $C$ using equation (\ref{CCc}), it is necessary to understand the characteristics of $R$, which in turn implies the need to comprehend the properties of $S$ and $\varGamma$. Once an observational array is established, $S$ and $\varGamma$ are typically fixed and can be simulated using electromagnetic simulation software or measured in a laboratory.

Calibrating this cross-talk effect from the response of a coherent interferometer is challenging, and knowledge of the S-parameter matrix will be key to calibration. Here, we simulate the $S_{\rm nn}$ parameters of the SIGMA array shown in Figure \ref{SNN}. From Figure \ref{SNN}, we can see that in the operating band of our greatest interest, i.e., 65-90 MHz, the $S_{\rm nn}$ values for antennas 4 through 17 are nearly identical, whereas antennas 1-3 and 18-20 exhibit more noticeable deviations. In practical observations, these edge antennas (1-3 and 18-20) will not be used. Therefore, we can assume that the S-parameter matrix under this array design has class properties of the topliez matrix. The magnitude and Phase plot of the simulated SIGMA S parameters matrix in 75 MHz is shown in Figure~\ref{fig:sm_combined}.

Our prior design approximated $S$ as a Hermitian Toeplitz matrix and ensured that the diagonal elements of $\varGamma$ are approximately equal, which implies that $R$ will also be approximately a Hermitian matrix. This implies that it can be numerically solved for its inverse or an approximate inverse, so $C$ can be calculated by equation (\ref{CCc}).

  \begin{figure*}
  \centering
  \includegraphics[scale=0.48]{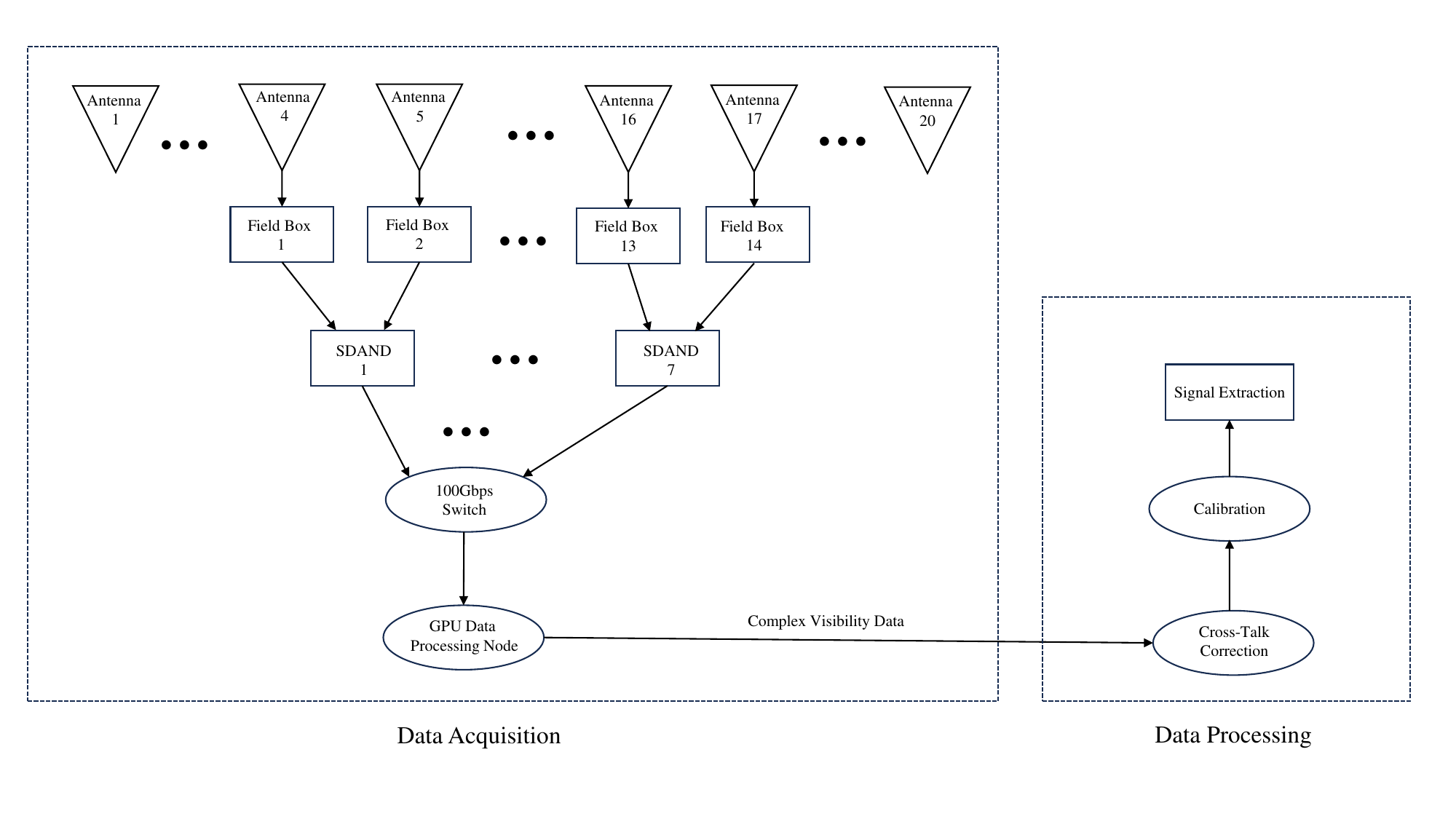}
  \caption{Block diagram of SIGMA system and data processing, antennas 4-17 will be used for data acquisition and other antennas are used as constraints, single antenna electronic system has been given in Figure \ref{fig:block}, SDAND represents the data acquisition node, each SDAND device processes two RF inputs, which are ultimately channelized and correlated at the data processing node to get the complex visibility data for subsequent data processing.}
  \label{sys}
  \end{figure*}

\section{Data Processing}
Here we discuss the process of collecting data and performing processing and extraction of global signals using SIGMA, a similar process has been proposed in \cite{mckinley2020all}, the block diagram of SIGMA system and data processing is shown in Figure \ref{sys} and in principle there are these steps in the pipeline:

\noindent\textbf{Data Acquisition} The radio-frequency signals from each antenna undergo low-noise amplification and band-pass filtering in a local field box before being passed via coaxial cables to digitization nodes. There are three low-noise amplifiers in the field box, specifically designed to minimize noise while boosting the signal strength. There is a customized 50-200 MHz band-pass filter that is applied between the amplifier stages and before analog-to-digital conversion. Once the signals are amplified and filtered, they are transmitted via coaxial cables to the digitization nodes. Each node is equipped with high-speed analog-to-digital converters (ADCs) that sample the incoming RF signals at a precise rate, converting the analog signals into digital data streams. Leveraging sub-nanosecond clock synchronization provided by the White Rabbit system, all nodes simultaneously sample the incoming signals in real-time. The digitized data streams from each node are then transferred via high-speed optical fibers to a remote GPU correlation center. This center is responsible for performing real-time channelization and correlation of the data. The channelization process involves dividing the frequency range into multiple spectral channels, allowing for the analysis of the signal across different frequencies. The correlation process combines signals from different antennas, calculating the complex visibility at each frequency and time step. This process generates cross-correlations between pairs of antennas and auto-correlations for individual antennas. The obtained correlated complex visibility data were used for subsequent analyses.

\noindent\textbf{Cross-Talk Correction} After obtaining the complex visibility, by using the known or measured system scattering parameters and the reflection coefficient of the low-noise amplifiers, we can construct a correction matrix to approximately remove the impact of cross-talk. In the absence of significant amplifier noise, the measured correlation matrix can be related to the ideal one, allowing us to invert or iteratively solve for the cleaner version of the correlation matrix. As mentioned in Section \ref{sect:ct}, once our observational array is established, its scattering matrix $S$, and the LNA's reflection coefficient matrix $\varGamma$, are typically fixed. $S$ and $\varGamma$ can be measured in a laboratory or be simulated using electromagnetic simulation software. Thus, we can obtain the matrix $R = (I - S \varGamma)^{-1}$. While disregarding the noise produced by the LNA, one could calculate the approximate $C$ from our observation data $\widetilde{C}$ according to equation (\ref{CCc}), to reconstruct the signal of the sky.

\noindent\textbf{Calibration} The complex visibility needs to be calibrated, and data from all baselines are used in the calibration process. The global sky model (\citealt{gsm}) can be used as a reference in this process. The GSM sky model is a global sky model constructed from all publicly available total power large-area radio surveys, covering frequencies from 10 MHz to 100 GHz, using a principal component-based fitting method. We can first generate GSM healpix sky maps at each frequency, and then create the reprojected map. The reprojected map is used as the model image and predicted into the MODEL column of the measurement set. Then we need to solve for the complex antenna gains using this input model. A variety of software is now available to solve antenna gain, such as CASA (\citealt{casa}), CALIBRATION (\citealt{calibration}), Hyperdrive \footnote{https://github.com/MWATelescope/mwa\_hyperdrive}, FHD (\citealt{fhd}). As an example, the calibration in hyperdrive is based on the sky model, data visibilities are compared to the MODEL visibilities, and the differences are used to calculate antenna gains. Here is the algorithm used to determine antenna gains:
\begin{equation}
    G_{p,i}=\frac{\sum_{q,q\ne p}{D_{pq}\,G_{q,i-1}\,M_{pq}^{H}}}{\sum_{q,q\ne p}{\bigl( M_{pq}\,G_{q,i-1}^{H} \bigr) \,\bigl( M_{pq}\,G_{q,i-1}^{H} \bigr) ^H}}
\end{equation}
Here $p$ and $q$ are antenna indices, $G_{p}$ is the gain Jones matrix for antenna $p$, $D_{pq}$ is a data Jones matrix from baseline $pq$, $M_{pq}$ is the model Jones matrix from baseline $pq$, $i$ is the current iteration index, and the superscript $H$ denotes the Hermitian transpose. This algorithm will be executed independently for each channel and once a calibration solution is obtained, the calibration solution will be applied to the input visibility data. After calibration, the data is converted to a format more suitable for further analysis.

\noindent\textbf{Signal Extraction} After obtaining the calibrated visibility data in the appropriate format, we will select the baselines among them that are sufficiently responsive to the global signal. The length of the baseline for each pair of visibility data is obtained by calculating $\sqrt{u^2+v^2}$, where only data shorter than 0.5 wavelengths will be used for subsequent analysis. The selected visibility data will then be converted to brightness temperature units (K) by multiplying the conversion factor $10^{-26}\lambda ^2/2k$, where $\lambda$ is the wavelength and $k$ is the Boltzmann constant.

Each set of measured visibility fits the description of equation~\ref{equationspr}, the real sky contains angular structures that will affect our data extraction. We can simulate the visibility of the angular structure of the sky based on the GSM map, beam model, and baseline, and subtract it from the observed visibility, in order to mitigate the effect of the angular structure. Then compute the global response based on the baseline and antenna radiation pattern model, and dividing it by the observed visibility gives the global sky temperature $T_{\rm sky}(\nu)$. This process is repeated across all frequency channels to generate a global signal spectrum. In principle, after obtaining the global signal spectrum, as \cite{bowman18}, we can use a simple polynomial foreground model to fit the data:
\begin{equation}
    T_{\mathrm{F}}(\nu )=\sum_{\mathrm{n}=0}^{\mathrm{N}-1}{a_{\mathrm{n}}\nu ^{\mathrm{n}-2.5}}
\end{equation}
where -2.5 is the exponent used to match the shape of the foreground, N is the polynomial order, $a_n$ is the fitting coefficient, and $\nu$ is the frequency. By fitting the data in log-log space and analyzing the residuals, or jointly fitting the CD/EoR signal model with the foreground model, we can obtain an estimate of the CD/EoR signal. The method described here is the basic principle of separating angular structures and extracting CD/EoR signals from short-spacing interferometer data, simple polynomials could be unreliable enough in practical data processing (\citealt{rao2017modeling}), but can be used for technical validation.

It should be noted that further quantitative analysis and simulations are needed to verify the calibration and signal extraction procedures described here fully. The detailed estimation of calibration errors and signal extraction will be addressed in future work.

\section{Discussions and Conclusions}
\label{sect:6}

In this paper, we present the design of the SIGMA, a novel interferometric system aimed at detecting the global 21cm signal from the CD and the EoR. The SIGMA experiment employs a unique approach by utilizing short-spacing interferometers. In principle, a potential advantage of short-spacing interferometers is that they may be more insensitive to uncorrelated receiver noise. This work thoroughly investigates the antenna design and the SIGMA layout.

After comparing different antennas, the resulting design adopts a parallel configuration 1-D array using blade antennas. We investigate the radiation pattern and the $S_{\rm nn}$ of the antenna array through simulations, where the mutual coupling effect when the antennas are closely spaced results in changing performance.

Another significant challenge we face is the cross-talk between antennas. Mutual excitation between antennas is expected to introduce numerous issues in subsequent data processing, necessitating further experimentation and testing. We discuss cross-talk caused by sky signals in antenna systems, noting that these effects can be calibrated once we have a comprehensive understanding of the entire system.

The SIGMA antenna layout showcases a 1-D linear array with 20 antennas, offering potential for future expansion. In the short term, our aim with SIGMA is to delve deeper into understanding the system's instrumental effects and to optimize its performance further.

Future work will focus on improving the antenna design and layout, continuing to explore mutual coupling, optimizing the receiving system, and continuing to monitor the experimental site for RF interference. The feasibility of calibration and signal extraction procedures, including quantitative error estimation and foreground removal simulations, will also be investigated in future work.

\section*{Data Availability}
The data underlying this article will be shared on reasonable request to the corresponding author.

\begin{acknowledgements}
This work is supported by National SKA Program of China No.SQ2020SKA0110200 and No.SQ2020SKA0110100.  YY acknowledges the support of the Key Program of National Natural Science Foundation of China (12433012).
\end{acknowledgements}

\bibliographystyle{raa}
\bibliography{ms2024-0376}

\label{lastpage}

\end{document}